\documentclass[sigplan,screen]{acmart}

\usepackage[utf8]{inputenc}
\usepackage[T1]{fontenc}

\usepackage{adjustbox}
\usepackage{booktabs}
\usepackage{listings}
\usepackage{mdwlist}
\usepackage{multicol}
\usepackage[autolanguage]{numprint}
\usepackage{proof}
\usepackage{softdev}
\usepackage{stmaryrd}
\usepackage{setspace}
\usepackage{subcaption}
\usepackage{wrapfig}
\usepackage{xspace}
\usepackage{menukeys}
\usepackage[normalem]{ulem}

\usepackage[ruled]{algorithm2e} 

\SetAlFnt{\small}
\SetAlCapFnt{\small}
\SetAlCapNameFnt{\small}
\SetAlCapHSkip{0pt}
\IncMargin{-\parindent}

\startPage{1}

\let\origthelstnumber\thelstnumber
\makeatletter
\newcommand*\Suppressnumber{%
  \lst@AddToHook{OnNewLine}{%
    \let\thelstnumber\relax%
     \advance\c@lstnumber-\@ne\relax%
    }%
}

\newcommand*\Reactivatenumber[1]{%
  \setcounter{lstnumber}{\numexpr#1-1\relax}
  \lst@AddToHook{OnNewLine}{%
   \let\thelstnumber\origthelstnumber%
   \refstepcounter{lstnumber}
  }%
}

\newcommand{\qtt}[1]{`\texttt{#1}'\xspace}

\newif\ifarxiv
\arxivtrue

\newcommand{\validalloverall}{96.8\%\xspace}

\newcommand{\validhistoverall}{89.4\%\xspace}

\newcommand{\validstackoverall}{95.6\%\xspace}

\newcommand{\validlinejavaphp}{79.9\%\xspace}

\newcommand{\totalinsertions}{5,100\xspace}
\newcommand{\breakdownallvalidsame}{65.8\%\xspace}

\newcommand{\breakdownallinvaliddiff}{2.3\%\xspace}
\newcommand{\breakdownallnovalid}{25.3\%\xspace}
\newcommand{\breakdownallnoerror}{0.8\%\xspace}

\newcommand{\breakdownallexclude}{88.1\%\xspace}

\newcommand{\corpusphpfiles}{365\xspace}
\newcommand{\corpusphploc}{174,147\xspace}
\newcommand{\corpusjavafiles}{6,556\xspace}
\newcommand{\corpusjavaloc}{1,888,190\xspace}
\newcommand{\corpusluafiles}{32\xspace}
\newcommand{\corpuslualoc}{13,990\xspace}
\newcommand{\corpussqlfiles}{1,042\xspace}
\newcommand{\corpussqltests}{27,525\xspace}
\newcommand{\corpussqlloc}{95,533\xspace}
\newcommand{\corpusphpexprs}{2,474\xspace}
\newcommand{\corpusphpfuncs}{518\xspace}
\newcommand{\corpusjavaexprs}{52,687\xspace}
\newcommand{\corpusjavafuncs}{6,743\xspace}
\newcommand{\corpusluaexprs}{4,476\xspace}
\newcommand{\corpusluafuncs}{230\xspace}
\newcommand{\corpussqlstmts}{61,065\xspace}

\lstnewenvironment{lstdefault}[1][]
  {
    \lstset{
        numbers=left,
        language=Python,
        keywordstyle=\color{red!70!black},
        commentstyle=\itshape\color{gray!90!black},
        stringstyle=\color{green!60!black},
        basicstyle=\linespread{0.9}\footnotesize\ttfamily,
        numberstyle=\tiny,
        keepspaces=true,
        breaklines=true,
        captionpos=b,
        abovecaptionskip=0.8em,
        columns=fullflexible,
        xleftmargin=10pt,
        aboveskip=1em,
        belowskip=0em,
        showstringspaces=false,
        literate={\$}{{\$}}1,
        escapeinside={{<!}{!>}},
        #1
    }
}{}

\lstdefinelanguage{EcoGrammar}
{
    alsoletter={:,\:=,|},
    morekeywords={:, =, | },
    keywordstyle=\color{red!70!black},
    morestring=[b]",
    stringstyle=\color{green!60!black},
    commentstyle=\itshape\color{gray!90!black},
    comment=[l]{//},
    morecomment=[s]{/*}{*/},
}

\lstnewenvironment{lsteco}[1][]
  {
    \lstset{
        language=EcoGrammar,
        basicstyle=\footnotesize\ttfamily,
        numberstyle=\tiny,
        xleftmargin=10pt,
        showstringspaces=false,
        #1
    }
}{}

\setcopyright{acmlicensed}
\acmPrice{15.00}
\acmDOI{10.1145/3357766.3359530}
\acmYear{2019}
\copyrightyear{2019}
\acmISBN{978-1-4503-6981-7/19/10}
\acmConference[SLE '19]{Proceedings of the 12th ACM SIGPLAN International Conference on Software Language Engineering}{October 20--22, 2019}{Athens, Greece}
\acmBooktitle{Proceedings of the 12th ACM SIGPLAN International Conference on Software Language Engineering (SLE '19), October 20--22, 2019, Athens, Greece}

\begin{document}
\title{Default Disambiguation for Online Parsers}

\author{Lukas Diekmann}
\affiliation{%
  \department{Software Development Team}
  \institution{King's College London}
  \country{United Kingdom}}
\author{Laurence Tratt}
\orcid{0000-0002-5258-3805}
\affiliation{%
  \department{Software Development Team}
  \institution{King's College London}
  \country{United Kingdom}
}
\thanks{Authors' URLs: %
    L.~Diekmann~\url{https://lukasdiekmann.com/},
    L.~Tratt~\url{https://tratt.net/laurie/}.
}

\begin{abstract}
Since composed grammars are often ambiguous, grammar composition requires a
mechanism for dealing with ambiguity: either ruling it out by using
delimiters (which are awkward to work with), or by using disambiguation operators to
filter a parse forest down to a single parse tree (where, in general, we cannot
be sure that we have covered all possible parse forests). In this paper, we
show that default disambiguation, which is inappropriate for
batch parsing, works well for online parsing, where it can be overridden by the user if
necessary. We extend language boxes -- a delimiter-based
algorithm atop incremental parsing -- in such a way that default disambiguation
can automatically insert, remove, or resize, language boxes, leading to
the \emph{automatic language boxes} algorithm. The
nature of the problem means that default disambiguation
cannot always match a user's intention. However, our experimental
evaluation shows that automatic language boxes behave acceptably in \validalloverall of
tests involving compositions of real-world programming languages.
\end{abstract}

\maketitle

\section{Introduction}

\begin{figure*}
    \vspace{1.2em}
    \includegraphics[width=1.00\textwidth]{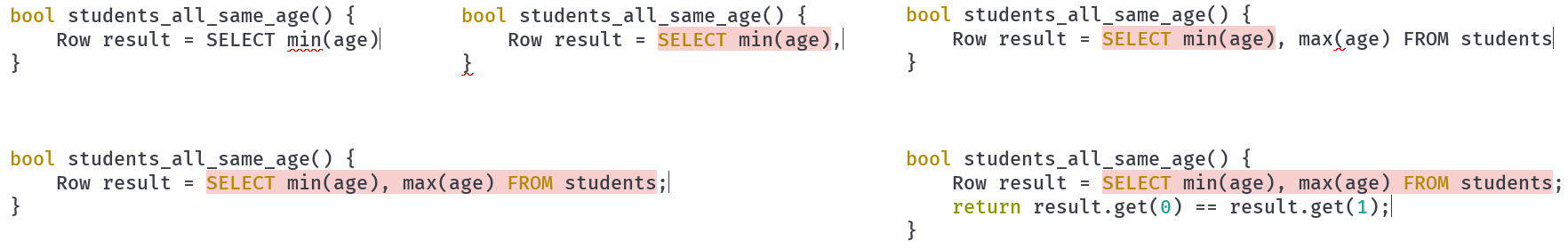}
    \begin{picture}(0,0)
        \put(-249,96){\textcolor{black}{\footnotesize\textbf{(a)}}}
        \put(-103,96){\textcolor{black}{\footnotesize\textbf{(b)}}}
        \put(  40,96){\textcolor{black}{\footnotesize\textbf{(c)}}}
        \put(-249,50){\textcolor{black}{\footnotesize\textbf{(d)}}}
        \put(  40,50){\textcolor{black}{\footnotesize\textbf{(e)}}}
    \end{picture}
    \vspace{-2.2em}
    \caption{An example of default disambiguation in action, with elided screenshots
      from our implementation of the automatic language boxes algorithm within
      the Eco editor. Here, the user is entering text in a
      composition of Java and SQL, where SQL statements can be used wherever
      Java expressions are valid. As the user types, language boxes (with a
      pink background) are automatically inserted, removed, or resized.
      \textbf{(a)} After typing the skeleton of a Java function, the user begins
      typing an SQL statement as the right-hand side expression of a Java
      assignment. The most fundamental part of the
      algorithm is to try inserting language boxes when a syntax error in the
      outer language occurs (as can be seen at the \texttt{min} function) but
      not if it then leads to a syntax error immediately after the inserted
      language box. It is thus too early to insert an automatic language box
      around the SQL as it would cause a syntax error in the first non-whitespace token
      afterwards (`\texttt{\}}'). \textbf{(b)} After typing `\texttt{,}' an SQL
      language box is automatically inserted since the first non-whitespace Java token
      (`\texttt{,}') is now syntactically valid. \textbf{(c)} The user continues
      typing a (now incomplete) SQL statement. This causes syntax errors in the
      outer language which cannot yet be resolved by inserting, removing, or
      resizing any language boxes.  \textbf{(d)} After typing `\texttt{;}', the
      automatic language box algorithm resizes the existing language box to
      encompass the entire SQL statement, making the program syntactically
      complete. \textbf{(e)} Further syntactically correct Java input does not
      cause the language box to be altered.
}
\label{intro_example}
\end{figure*}

Language composition -- the ability to build larger languages out of multiple
small languages -- offers an enticing solution to problems such as the
development of domain-specific languages or the migration of legacy software.
Unfortunately, writing and editing composed programs is often cumbersome.
Arguably the major cause of this is due to ambiguity when parsing: language composition is underpinned by grammar
composition; composing even two provably unambiguous grammars can lead to an
ambiguous grammar; and determining whether a grammar is unambiguous or not is
undecidable~\cite{cantor62ambiguity}. There are two fundamental approaches
to dealing with such ambiguity:~either ruling it out through the use of
delimiters (via explicit bracketing or syntax-directed
editing); or using a generalised parsing algorithm that can create a parse
forest, capturing all ambiguities
(e.g.~\cite{visser97scannerless}). Each has different trade-offs: delimiters
are visually intrusive and/or awkward to work with; and one can never know if
enough disambiguation operators have been used to filter all possible parse
forests down to a single parse tree.

In this paper we show that default disambiguation -- i.e.~a disambiguation
strategy applied equally to every language composition -- provides a satisfying new
point in the design space. This may seem surprising,
since default disambiguation is clearly inappropriate for batch
parsing, where a single incorrect disambiguation will cause dire results.
The realisation underlying this paper is that this is not an issue for online parsers,
since users can manually override incorrect disambiguations. However, this
introduces a new challenge: default disambiguation must mostly match the users
intentions, because if they have to override it too often then the system will
be perceived as unusable.

In this paper we use \emph{language boxes} as the basis of a default
disambiguation system. Language boxes aim to
combine the advantages of explicit bracketing and syntax-based
editing~\cite{diekmann14eco}. Editors
which support language boxes need to use incremental parsing
(we assume use of the~\citet{wagner98practicalalgorithms} algorithm):
unlike traditional editors, users do not directly edit
a contiguous block of text in memory, but instead
indirectly edit a parse tree, which is continuously updated as they type.
A language box is then simply a node in the parse
tree that represents a different language, surrounded by
explicit, but invisible, delimiters. Unlike explicit bracketing approaches,
there are no visually intrusive delimiters; unlike traditional
syntax-based editors, the program can be syntactically incorrect in arbitrary
ways and places during editing. Language boxes have the virtue that they work for any possible
language composition. However, this generality comes at a cost:~users
must explicitly, and tediously, state when they want
to insert or remove language boxes.

We thus created default disambiguation rules which can
automatically insert, remove, and resize language boxes in many useful cases
--- leading to the \emph{automatic language boxes} algorithm.
Given several languages in
a composition, automatic language boxes find a non-strict subset of the
possible ambiguous parses: in many cases, only one possibility remains, and
language boxes are inserted or updated as appropriate. The
algorithm is comprised of several stages / heuristics: determining what user
inputs should trigger it; finding candidate language boxes to insert, remove,
or resize; filtering out those which would make the overall program worse; and
then applying the remaining candidates. Figure~\ref{intro_example} shows a
simple example and walks readers through the high-level parts of the algorithm
and accompanying heuristics.

We implemented automatic language boxes as an extension to the Eco
editor~\cite{diekmann14eco}. In order to validate automatic language boxes, we
created 12 language compositions involving large, real-world languages
(Java, Lua, PHP, and SQL),
and composed programs in those compositions by extracting fragments from real-world programs.
In essence, our experiment is equivalent to opening a file
in the outer language, moving the cursor to a given position, deleting a fixed amount
of text, and then inserting text one character at a time from the inner language.
We then group the possible outcomes of such actions into two overall categories: `acceptable' (roughly
speaking: a language box was inserted without causing an error; or no language
box was inserted because the fragment is also valid in the outer language) and
`unacceptable' (a language box was inserted but caused an error elsewhere in
the file; or no language box was inserted despite the fragment being invalid in
the outer language). Across the \totalinsertions tests we ran, \validalloverall are
classified as acceptable, with the majority
(\breakdownallvalidsame of the total) leading to a language box being inserted
around the entire fragment, and most of the rest (\breakdownallnovalid of the total)
being instances of the fragment being valid in both the outer and inner
languages. We believe that this data shows that default disambiguation is
a practical means for editing composed programs. Our fully
repeatable experiment can be found at \url{https://archive.org/details/defaultdisambiguation}.

\section{Background}
\label{sec_background}

In this section, we briefly survey existing approaches to editing composed
programs, before giving a brief overview of incremental parsing, sufficient for
this paper's purposes.

\subsection{Delimiter-based Approaches}

The traditional approach to language composition is to use delimiters
between languages. The
most obvious way of achieving this is to use explicit brackets to make clear a
switch from an outer to an inner language (e.g.~`\texttt{for (String e: <<SELECT name
FROM table>>) \{ ... \}}'), though this is visually intrusive
(what~\cite[p.~4]{bravenboer05generalized} call ``syntactic clutter``), and
prevents the
brackets being used within the sub-language (e.g.~in this case, the inner language
cannot use `\texttt{>>}' as a bit-wise operator).

Naive approaches inherit a severe restriction from traditional parsing, which
separates lexing (i.e.~the splitting of the user's input into tokens) from
parsing (i.e.~the structuring of tokens into a parse tree): all the languages
in the composition
must share the same lexing rules. This restriction can be somewhat eased if the
lexer recognises the explicit brackets and extracts text between them wholesale
for separate lexing and parsing (see e.g.~\cite[p.~13-14]{tratt08domainspecific}),
though it is then hard for the lexer to accurately keep track of nested
brackets (e.g.~should brackets in comments be counted or not? and how does
one know what format comments in the inner language(s) are in?).  A
more sophisticated approach is for the lexer and parser to interact (see
e.g.~\cite{wyk07context}), such that the parse causes a switch in lexing rules
when input shifts to an inner language.
This has the advantage that brackets do not always need to be quite as visually
intrusive (e.g.~one can use a difference in keywords to identify a switch from
one language to another), though in the general case explicit brackets must still be
used to resolve ambiguities.

\subsection{Scannerless Parsing}
\label{sec:scannerless}

Generalised parsing can parse any Context-Free Grammar (CFG), even those that
are ambiguous. Scannerless parsing~\cite{visser97scannerless} extends this such
that lexing and parsing are specified together. This removes the need for
explicit brackets entirely, but leads to more ambiguities, since
traditional lexers resolve many ambiguities (e.g.~between identifiers and keywords)
before parsing. This is challenging because ambiguity is, in general,
undecidable~\cite{cantor62ambiguity} and even the best ambiguity heuristics fail
to find all possible sources of ambiguity~\cite{vasudevan13detecting}. Thus,
no matter how many static disambiguation operators one uses, in general one
cannot be sure if all possible points of ambiguity have been covered.
Furthermore, disambiguation operators can cause scannerless
parsers to become context-sensitive~\cite{eijck07accept}, the
consequences of which remain unclear. Although it is possible in some cases
to use semantic information such as types to aid disambiguation (see
e.g.~\cite{vinju05typedriven}), this is not applicable to all languages.

\subsection{Syntax Directed Editing}

Traditional syntax directed editing avoids parsing text entirely. In essence,
users edit an AST directly, with incomplete parts of a program being
represented by holes. This avoids the need for explicit delimiters, and
sidesteps issues of ambiguity completely. However, such systems are awkward to
use~\cite[p.~2]{khwaja93syntax}, for example only allowing complete subtrees in the
AST to be selected at a time (e.g.~for the expression `2 + 3 * 4' one can
select `2' or `3 * 4', but not `2 + 3'),
and quickly fell out of fashion. The modern syntax directed editor
MPS~\cite{pech13mps} alleviates some, though not all, of these problems.
However, it requires significant expertise on the part of the language composition
author to make editing a pleasant experience, as the AST structure places
constraints on many editing operations.

\subsection{Incremental Parsing}

Parsing is traditionally a batch process: an entire file is fed through a parser
and a parse tree created from it. Incremental parsing, in contrast,
is an online process, continually parsing text and updating a parse tree as the user types.
In this paper we make use of the incremental lexing and
LR incremental parsing algorithms of~\citet{wagner98practicalalgorithms},
taking into account the several fixes found in~\citet{diekmann18editing}.
In this subsection we provide a brief overview of this algorithm sufficient
to understand the rest of this paper.

The incremental lexer and parser both operate on the parse tree. Parse tree nodes
are either \emph{nonterminals} (representing rules in the grammar) or
\emph{tokens} (representing terminal symbols). Nonterminal nodes have an
immutable type (e.g.~`expr') but a mutable list of child nodes.
Tokens have a mutable type (e.g.~`int')
and a mutable value (e.g.~`3') but no children.

After input from the user is received, the incremental lexer is run first.
Using lookahead information, it identifies the affected area of the change,
updates or creates tokens as necessary, and marks the path from each updated or
created token to the root as changed. The incremental parser then runs,
reparsing all subtrees with changes in them, and creating or removing
nonterminals as needed.

\subsection{Language Boxes}

\begin{figure}[t]
\begin{center}
\includegraphics[width=0.35\textwidth]{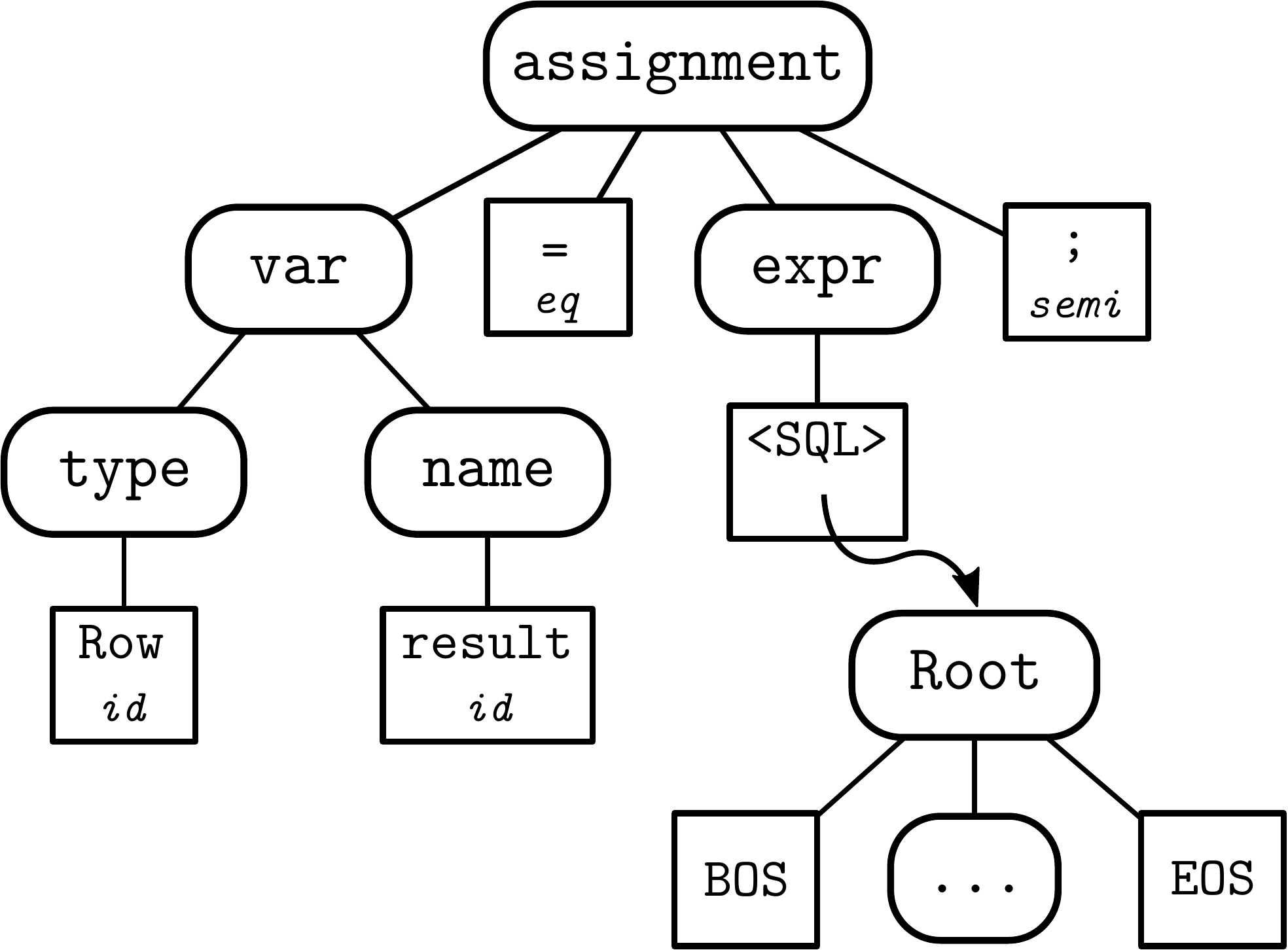}
\caption{An elided example of a parse tree in an incremental parser with
  language boxes: nonterminals have a type and zero or more children; terminals
  have a type (top) and a value (bottom). The composition in question is,
  again, (outer) Java and (inner) SQL. Here, the outer Java code is an
  assignment (`\texttt{type name = ...;}'). The right-hand side of the
  assignment is an SQL language box (the node with type `\texttt{<SQL>}'): from the
  perspective of the outer Java code, the SQL node is a terminal (and hence its
  value is irrelevant). In reality, the SQL node has a complete SQL parse tree
  underneath it: the special Root, BOS (Beginning Of String), and EOS (End Of
  String) nodes that every incremental parse tree contains, as well as the actual
  SQL contents (elided to `...' in this example). }
\label{fig:lboxtree}
\end{center}
\end{figure}

Language boxes allow users to embed one language inside another in the
context of an incremental parser. From the perspective of an outer language, a
language box is a terminal. Since
parsers care only about the type of a terminal, this is a natural fit. In
reality, language boxes do have content, though it is not visible to the outer
language: they contain a separate parse tree for the inner language within
them.

This simple definition belies its power. Consider our running Java and
SQL example composition. Java's grammar must have a reference from Java's
expression rule to a special symbol type `language box' (conventionally
represented between angle brackets to visually separate it from rules and
tokens). At run-time, if the Java parse tree has an SQL language box at the
correct point, then Java considers the tree to be syntactically correct. The
SQL language box will have its own SQL parse tree inside which may or may not
be syntactically correct. An elided example of such a tree-of-trees can be seen
in Figure~\ref{fig:lboxtree}.

Philosophically, language boxes thus form delimiters, albeit invisible
ones, between languages: from a syntactic perspective, outer languages are
ignorant of the contents of inner languages and vice versa. Thus we get much of
the power of syntax-directed editing without the accompanying difficulty of
editing ASTs. At all points, all languages can be manipulated as normal text
using the incremental parsers.

\section{The Outlines of a Solution}

\label{the problem}
The chief weakness of language boxes is that they must be inserted
manually --- this involves pressing a special key combination,
selecting the desired inner language from a list, typing the content, and (in
general) pressing a second special key combination to complete the language box.
When language boxes are used infrequently, this is merely irritating,
but when language boxes are used frequently, it is a significant usability
issue, impeding the user's flow. Eco, an editor which supports language boxes~\cite{diekmann14eco},
slightly eases this problem by using information from the incremental parser to
highlight those languages valid at the point of the cursor,
though this is a mild palliative at best.

In an ideal world, we would be able to automatically insert language boxes exactly, and only,
when they are wanted by the user. However, this is impossible in the
general case because language composition is really grammar composition
in disguise, and thus subject to the same ambiguity problems as generalised
parsing (see Section~\ref{sec:scannerless}). We must thus lower our sights
slightly, aiming to automatically insert, remove, or resize language boxes
correctly merely in the vast majority of cases, with the remaining cases
few enough in number that the user is willing to override them manually.
This still leaves a broad solution space which we narrow down with the
following soft considerations.

First, a solution which requires language composition authors (i.e.~the people
who actually compose grammars, create code generators etc.) to provide additional
hints or commands to aid automatic language box insertion is less likely
to be used widely and/or correctly. The meta-system underlying a language
composition system is often complex, and expecting language composition authors
to be expert in every part of it (as well as the domain they are composing
languages for!) is unrealistic. For example, it can be difficult to know whether
a non-LR grammar is ambiguous or not~\cite{vasudevan13detecting} and whether
one has disambiguated it in the expected way: grammar
composition magnifies such concerns, particularly as the number of languages
in a composition grows.

Second, a solution which seriously degrades performance would be unacceptable.
For example, one simple way of finding which language boxes to insert would
be to reparse the complete file on every keypress, which would be noticeably slow for large
files. Ideally, the theoretical performance guarantees
of~\citet{wagner98practicalalgorithms} would be maintained as well as good
practical performance\footnote{Interestingly, the original implementation of
this incremental parsing algorithm had to be triggered by the user (e.g.~when
a file was saved). Modern machines are fast enough that even a naive
implementation can run comfortably on every keypress in nearly all reasonable
cases.}.

Third, a solution which inserts language boxes unpredictably is unlikely
to find favour with users. Clearly, given the hard constraint described at the
start of this subsection, users cannot expect perfect language box insertion
all of the time. However, a reasonable minimum expectation is that it should be
entirely predictable as to when automatic language boxes are potentially
inserted, removed, or
resized; and, ideally, largely predictable as to what the effects of such
actions are. Furthermore, false negatives (i.e.~when the system inserts, removes,
or resizes language boxes incorrectly) are likely to be particularly harshly
received by users and must be reduced to the minimum possible.

\section{Automatic Language Boxes}

In this section, we present a default disambiguation mechanism in the form of
the \emph{automatic language boxes} algorithm. Given an arbitrary language
composition, it uses several heuristics to find plausible places to insert,
remove, or resize language boxes. The algorithm makes use of the fact that it
has a surrounding parse tree to provide context, and knowledge of where
the user has recently made edits, to improve the quality of its results.

To ease the algorithm's description, we start by considering the problem of language box
insertion, before then adding additional functionality (e.g.~removal and
resizing). We later validate the usefulness of automatic language boxes in
Section~\ref{sec:evaluation}.

\subsection{The Consideration Heuristic}

The first challenge with automatic language boxes is to decide upon a sensible
heuristic for considering if/when to insert a language box -- what we call the
\emph{consideration heuristic}. If the consideration heuristic triggers too
frequently, it will lead to too many unwanted language boxes being inserted,
each of which must then be manually removed by the user.
Conversely, if it triggers infrequently, it will not be a useful aid to the user.

We use two related observations as the guides to our consideration heuristic.
First, by definition, language composition always consists of an outer language
and one or more inner languages\footnote{Note that these terms are relative:
when we create a language box and move into it, the previously inner language
now becomes the outer language.}. It is thus a reasonable expectation that most
text typed in the outer language is intended to be in the outer language.
Second, the clearest indication that recently typed text in the outer language
might have been intended for an inner language is that it leads to a syntax error
in the outer language.

\begin{figure}[tb]
    \includegraphics[width=.40\textwidth]{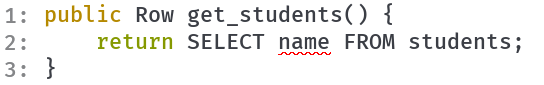}
    \caption{An example of a syntax error in a Java and SQL composition. In
      this example, we have turned off automatic language box insertion to
      emphasise the fact that syntax errors often occur in the middle of the
      language box we would like to insert.}
\label{fig:consideration}
\end{figure}

Our consideration heuristic therefore triggers at the point of each new syntax error.
This is an entirely predictable heuristic from a user perspective, though it
does have two consequences: the point of a syntax error is not always at the
beginning or end of the text that a user expects to be put in a language box
(see Figure~\ref{fig:consideration}); and this heuristic clearly works better
for languages whose syntaxes don't overlap a great deal (see
Section~\ref{sec:highly ambiguous compositions}). Happily, we can rely
on the fact that the incremental parser isolates syntax errors after they
occur~\cite[p.~93]{wagner98practicalalgorithms}, so that there is no
possibility of old syntax errors being considered a second time.

\subsection{The Candidates Heuristic}

Once the consideration heuristic has triggered, we then have to search for
plausible nodes in the parse tree at which
language boxes could be inserted -- what we call the
\emph{candidates heuristic}. The trade-off here is that identifying too many
locations slows down the search and overwhelms the user with possibilities; but
identifying too few locations means that useful candidates are missed. We thus
define several sub-heuristics which we then combine together. A candidates heuristic
can produce zero or more candidate language boxes at any given point; those
candidate language boxes may cover different spans and/or be of different
language types. Before we define the candidates heuristics themselves, we first
introduce the concept of recognisers, which all candidates heuristics use.

\subsubsection{Recognisers}
\label{sec:recognisers}

\begin{figure}
\begin{lstdefault}[]
def cnds_recogniser(node, lang):
  lexer = <!\textrm{\textit{lexer for lang starting at node}}!>
  parser = <!\textrm{\textit{parser for lang}}!>
  cnds = []
  while True:
    token = lexer.next_token()
    if token is None:
      return cnds
    parser.parse_token(token)
    if parser.accepted():
      cnds.append((node, token.end_pos, lang))
    elif parser.error_node.type_ != "EOS":
        return cnds
\end{lstdefault}
\caption{A generic candidates recogniser which produces the ending offsets of
  each substring starting at \texttt{node} that is valid in language
  \texttt{lang}. In essence, we create a lexer and parser for \texttt{lang}
  (lines 2--3) and then try recognising substrings that grow one token at a
  time (lines 6 and 9), though note that the recogniser parser reuses the previous
  state. If we reach the end of the parse tree we are complete
  (lines 7--8). If we successfully parse a substring, we add a candidate to the
  list (lines 10--11). If a substring causes a parse error on anything but the
  implicit EOS (End Of String) token, we know that growing the substring
  further cannot fix the parse
    and terminate the search (lines 12--13).}
\label{fig:recogniser}
\end{figure}

When a candidates heuristic has identified a node $n$ in the parse tree as the
plausible start of a language box, we then have to decide if one or more
language boxes could start at that point. Although it would be possible to use
the normal incremental parser to answer this question, it would require
significant setting up and tearing down which would be tedious to program and
slow to run. Instead we provide \emph{candidates recognisers}\footnote{Some languages
(e.g.~whitespace sensitive languages such as Python) need slightly customised
candidates recognisers.} which are able to quickly return the list of substrings valid in
a language $L$ starting at node $n$.

\begin{figure*}[t]
\begin{minipage}[t]{0.5\textwidth}
\begin{lstdefault}[]
def parse_tree(parser, node):
  cnds = []
  v = global_version - 1
  while node is not None:
    for lang in <!\textrm{\textit{composition}}!>:
      if <!\textrm{\textit{lang can be shifted before \texttt{node.version(v)}}}!>:
        cnds.extend(cnds_recogniser(node, lang))
    node = node.parent(v)
  return cnds

def stack(parser, node):
  cnds = []
  for state, node in reversed(parser.stack):
    for lang in <!\textrm{\textit{composition}}!>:
      if <!\textrm{\textit{lang can be shifted at \texttt{state}}}!>:
        t = node.next_terminal()
        cnds.extend(cnds_recogniser(t, lang))
  return cnds
\end{lstdefault}
\end{minipage}
\begin{minipage}[t]{0.48\textwidth}
\begin{lstdefault}[firstnumber=19]
def line(parser, node):
  cnds = []
  while node.type_ not in ["BOS", "Newline"]:
    for lang in <!\textrm{\textit{composition}}!>:
      if <!\textrm{\textit{lang can be shifted before \texttt{node}}}!>:
        cnds.extend(cnds_recogniser(node, lang))
    node = node.prev_terminal()
  return cnds
\end{lstdefault}
\end{minipage}

  \caption{Simplified versions of our three candidates heuristics. Each is
  passed a \texttt{node}, which is the point at which an error is detected, and
finds sensible points before that node in the parse tree to be the possible
starting point of language boxes (using the candidates recogniser from
Figure~\ref{fig:recogniser}).
The \texttt{parse\_tree} candidates heuristic
walks up the parse tree to find candidate language boxes (line 6--7).
Because the parse tree is, in a sense, only partially parsed at the point the
candidates heuristic is called, we have to view the parse tree as it was before parsing
began (line 8).
The \texttt{stack} candidates heuristic walks the parse
stack, finding the node matching each point in the stack (line 13) and then
searches for candidate language boxes (line 15) starting at the first terminal
following that node (line 16).
The \texttt{line} candidates heuristic searches
backwards each token from the error node (line 25) until the beginning of
the line which contains that node (line 21) for
candidate language boxes (line 23).}
\label{lst:find_candidates}
\end{figure*}

The main challenge for candidates recognisers is to decide when to stop trying to recognise
further input. If we stop too early, we will fail to recognise valid language
boxes, but if we go too far, we will degrade performance. The technique
we use is to try recognising gradually growing substrings as valid in an inner
language, making use of the fact that the recogniser parser implicitly
reuses state from the previous token.
If a substring is not valid, we then check where the parse failed:
if it failed on the EOS (End Of String) token, then it is possible that
extending the substring might lead to a valid parse, so we continue the search; but if it
failed earlier than the EOS token, then we know that extending the substring
cannot fix the parse and we stop the search. Figure~\ref{fig:recogniser} shows a
more formal version of this algorithm.

For example, consider the fragment `\texttt{int x = SELECT 1 + 2;}' in our
running Java and SQL composition. If we start a candidates recogniser at the
`\texttt{SELECT}' token, we first try recognising `\texttt{SELECT}' which leads
to a syntax error at the EOS token, so we continue. We then try recognising
`\texttt{SELECT 1}', which is valid SQL, so we add it to our candidates list
and continue. `\texttt{SELECT 1 + }' errors at the EOS token, so we continue.
`\texttt{SELECT 1 + 2}' succeeds, so we add it to our candidates list.
`\texttt{SELECT 1 + 2;}' errors at the `\texttt{;}' token, so the search then
terminates, even if there is input after the fragment.

\subsubsection{The Parse Tree, Stack, and Line Heuristics}

We eventually created three distinct candidates heuristics, each of which has
different strengths and weaknesses. We now describe each candidates heuristic
in detail; see Figure~\ref{lst:find_candidates} for a semi-formal version of
each.

The \emph{parse tree} candidates heuristic aims to
find plausible candidates based on the structure of the parse tree. The
intuition underlying this is that a likely point to insert a language box is
around text that forms a subtree and that we can find such points by
recursively walking the parent nodes of the node in which a syntax error was
found. However, there is a slight subtlety in that the parse tree is, by
definition, broken at the point the candidates heuristic is called. In a sense,
the incremental parser parses the tree in two stages, and the candidates
heuristic is called after the first of these, when it is possible for newly
inserted terminals to be detached from the tree~\cite[p.~58, 60]{wagner98practicalalgorithms}. Fortunately, we can solve this
easily by using the versioning feature described in ~\citet[p.~15]{wagner98practicalalgorithms}
which allows
us to view the tree as if the first stage of parsing had not yet occurred.

The \emph{stack} candidates heuristic is based on
the idea that each point in the parsing stack naturally defines a plausible
breaking point between one language and another. It walks backwards over the
parsing stack, at each point looking at the associated node. This heuristic has
two significant
advantages: the parsing stack is nearly always small, so few additional
places in the program need to be checked; and if a language box can be
inserted, parsing can continue as normal from that position in the parsing
stack.

The weakness of the parse tree and stack heuristics is that both try relatively
few locations, and that they rely on the structure of the underlying LR
grammars, which does not always match human expectations of a language's
structure. The \emph{line} candidates heuristic is therefore very different,
and captures the intuition that many language
boxes are intra-line: it searches
backwards for candidates, one node at a time, from the error node to the beginning of the line
that contains the error. This heuristic
ensures that all
candidate locations close to the error node are searched, but bounds the search
in a way that is unlikely to cause a noticeable slowdown.

In order for later stages in the algorithm to work correctly, each candidate
needs to have a valid parsing stack. While candidates produced by the stack
heuristic naturally do so, we need to create a
valid parsing stack for candidates from the parse tree and line heuristics.
Fortunately, we can do this efficiently by
using a similar approach to that used by the incremental parser to reparse
nodes (see Figure~\ref{fig:createparsestack}).

\begin{figure}
\begin{lstdefault}[]
def recreate_parsing_stack(lbox):
  v = global_version - 1
  path_to_lbox = set()
  parent = lbox.parent(v)
  while True:
    path_to_lbox.add(parent)
    parent = parent.parent(v)
    if parent is None:
      break

  parser = <!\textrm{\textit{initialise parser}} !>
  node = <!\textrm{\textit{root node}} !>
  while node is not lbox:
    if node in path_to_lbox:
      node = node.children(v)[0]
    else:
      parser.parse(node)
      node = node.next_lookahead()
\end{lstdefault}
\caption{An algorithm for efficiently creating a parse stack after a language box
has been inserted into the parse tree. We do this in two steps. First we
collect all the nodes on the path from the language box to the root node (lines
2--9). Second we then follow equivalent steps as when a node is marked as
changed and the incremental parser runs \cite[p.~63]{wagner98practicalalgorithms}:
we reparse all nodes up to the language box (lines 11--18), skipping
subtrees which can't be relevant to the parsing stack (lines 17--18). The
\texttt{next\_lookahead} function (line 18) returns the next node in the parse
tree in preorder. Note that we have to iterate over a previous version of the
parse tree: \texttt{node.parent(V)} and \texttt{node.children(V)} both work in
the same way e.g.~ \texttt{node.parent(\textit{V})} returns \texttt{node}'s
parent as when \texttt{node} was in version \texttt{V} (which may be
different to \texttt{node}'s parent in the current version of the tree,
\texttt{global\_version}).}
\label{fig:createparsestack}
\end{figure}

\subsection{Combining Heuristics}
\label{sec:allheuristic}

As we shall see in Section~\ref{sec:evaluation}, each of our candidates
heuristics has strengths and weaknesses. We therefore combine them into a
single candidates heuristic, imaginatively called \emph{all}. This aggregates
the candidates from the individual heuristics, filters out those which
immediately break the surrounding context, ranks the remainder, and filters out all
of the non-best candidates.

We first filter out language boxes which are not valid in their immediate
context. By definition, our candidates heuristics will not have suggested
candidate boxes which are syntactically incorrect relative to the preceding
context, but they do not check the following context. A simple solution here
might seem to be to filter out language boxes which are followed by errors, but
this would lead to us filtering out candidates when the user has
deliberately left a later part of the program in a syntactically incomplete
state. We thus filter out only those candidates where the first non-whitespace
token following the candidate language box contains an error. The reason we
specify the first non-whitespace token, rather than simply the first token, is
that grammars for incremental parsing almost always define whitespace as a
token. This means that the incremental parser often inserts a whitespace token
after a candidate language box, and that whitespace token is by definition
syntactically valid, though not particularly insightful. We thus need to skip
such tokens in order to get to a token which tells us something useful about
the context surrounding a candidate language box.

We then rank candidates by how far we can parse successfully after them without encountering
a syntax error. However, since different language boxes can encompass different
parts of the input, and since inserting different language boxes can change how
much input subsequent tokens consume, finding a simple definition of ``how
far'' is surprisingly difficult. Our high-level solution is to continue
parsing after each candidate language box and use the absolute character
offset of the first subsequent parsing error as
a reasonable proxy for ``how far''. In order to bound this check, we first take
all the candidates and non-deterministically select one of those whose end character offset is the
equal largest: we then parse 10 further tokens beyond that language box's end
to find the `maximal parse point'. For each remaining candidate language
box, we then try parsing beyond it. If we hit a parsing error before the
maximal parse point, we discard the candidate language box; if we parse up to, or beyond,
the maximal parse point, we consider the candidate language box equally
good to the initial candidate language box. Note that the `or beyond' clause is needed
because different candidate language boxes may lead to the remaining input
being lexed in different ways, which may lead to tokens of different lengths
being created.

\subsection{Applying or Presenting Candidates}
\label{applying and presenting}

Once the `all' candidates heuristic has run, we will have zero or more possible
language boxes to insert. If there are zero candidates, then the algorithm
completes. If there is one candidate, we simply insert it to the user's
program. If the user is unhappy with the insertion, they can remove it by
pressing undo (conventionally \keys{Ctrl+Z}). However, implemented naively,
the language box can simply reappear on each subsequent keypress, which is
unlikely to be desired. Therefore removing a language box in this way marks
a flag \texttt{noinsert} on the error node identified by the
consideration heuristic. The consideration heuristic ignores nodes where
this flag is set to true so that language boxes are not
reinserted at a point where the user has explicitly indicated they do not
want them.

\begin{figure*}
\vspace{0.55em}
\begin{center}
\includegraphics[width=1\textwidth]{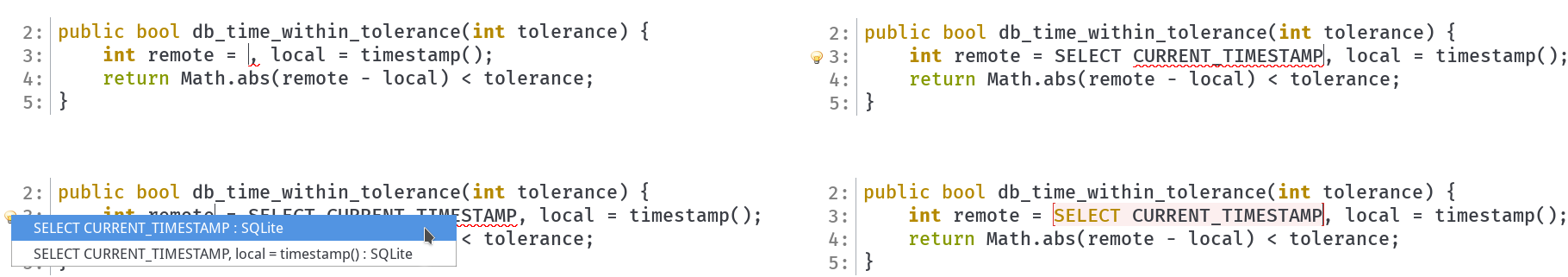}
\end{center}
\begin{picture}(0,0)
    \put(-247,102){\textcolor{black}{\footnotesize\textbf{(a)}}}
    \put(13,102){\textcolor{black}{\footnotesize\textbf{(b)}}}
    \put(-247,50){\textcolor{black}{\footnotesize\textbf{(c)}}}
    \put(13,50){\textcolor{black}{\footnotesize\textbf{(d)}}}
\end{picture}
\vspace{-2.0em}
\caption{An example of multiple language box candidates in our running Java and SQL
  composition. \textbf{(a)} The user is editing an existing Java programme,
  and has just deleted the expression after \qtt{remote}.
  \textbf{(b)} Inserting an SQL statement leads to a syntax error in Java. The
  automatic language box algorithm then finds multiple valid language box
  candidates. Rather than picking one at random, the syntax error remains, and
  the existence of multiple candidates is indicated to the user by the light
  bulb icon next to line 3.
  \textbf{(c)} Clicking on the light bulb displays the candidate language boxes
  that could be inserted. In this example, the user clicks on the first in the
  drop-down list. \textbf{(d)} The appropriate language box is inserted.}
\label{fig:multiplecnds}
\end{figure*}

\label{multiple candidates}
However, if there are multiple candidates, we then have two choices: we could
insert one of the candidates and present the others to the user as choices; or
simply present all the candidates as choices without inserting
any of them. The former approach is surprisingly hard to do well. If we were to
non-deterministically select one candidate and insert it, the user would be
unable to predict what was about to happen on each key press. We could instead
rank candidates (perhaps by `relevance', or length, or starting position etc.),
but we were unable to find a ranking system which matches the user's intentions
often enough to be worthwhile. We therefore simply present all the candidates
as options to the user from which they must choose one (see Figure~\ref{fig:multiplecnds}). As we shall see in
Section~\ref{sec:evaluation}, this happens rarely enough that it is not a
significant problem. It is also worth noting that this is an
example of a fundamental difference between batch and online parsing:
it is entirely feasible for us
to ask the user for their help in choosing language boxes as that choice can
be made once and recorded permanently, rather than having to be made anew
on each (batch) parse.

\begin{center}
\end{center}

\subsection{Removing and Resizing}

Automatically inserted language boxes start life in the \emph{uncommitted}
state, which means that they can then be considered possible candidates for
automatic removing and resizing. Language boxes move to the \emph{committed}
state when the user shows that they have finished
editing at the current point by moving the cursor outside of the language box.
Users can manually change a committed box to uncommitted if they
later want it to be subject to the algorithm again. If the content of an
uncommitted language box, or its surrounding area, changes then it may be
automatically removed or resized by the algorithm.

\subsubsection{Removing}
\label{sec:removing}

\begin{figure*}
\begin{center}
\includegraphics[width=0.70\textwidth]{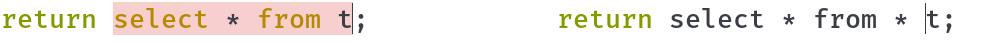}
\begin{picture}(0,0)
    \put(-375,5.5){\textcolor{black}{\textbf{(a)}}}
    \put(-175,5.5){\textcolor{black}{\textbf{(b)}}}
\end{picture}
\vspace{-1.1em}
\end{center}
\caption{A (slightly contrived) example showing automatic language box removal
  in our Java and SQL composition.
\textbf{(a)} The user meant to write a Java expression, but forgot
the second \qtt{*} after \qtt{from}: thus an SQL language box
was inserted.
\textbf{(b)} The user moves the cursor to before \qtt{from} and inserts the
missing \qtt{*}. This makes the contents of the SQL language box syntactically
invalid. Since its contents are, however, valid in the outer language (Java),
the language box is removed.}
\label{fig_autoremoval}
\end{figure*}

The simplest example of when we might want an automatic language box to be removed is if
the user deletes a character immediately after an automatic language box has
been inserted. For example, if an SQL language box is inserted as soon as the
user types `\texttt{int x = SELECT * FROM t}' and the user then presses
backspace, the automatic language box should disappear because `\texttt{SELECT
* FROM}' is valid Java and we want to prioritise the outer language in a
composition (see Figure~\ref{fig_autoremoval}). The full set of situations that
we handle is as follows:

\begin{enumerate}
  \item If an uncommitted language box has a syntax error within it, and
    if its contents are valid in the outer language, then the language box
    is removed.
  \item If an uncommitted language box becomes part of a syntax error in the outer
    language, and if its contents are valid in the outer language, then the
    language box is removed.
  \item If an uncommitted language box's content is valid in both the inner and
    outer languages, then the language box is removed if that would not
    then cause subsequent parse errors. Following the precedent from
    Section~\ref{sec:allheuristic}, we use the first non-whitespace token after
    the language box as a proxy for this.
\end{enumerate}

Note, that while the first two situations are triggered by an error occurring either inside
the box or on the box itself, the third situation has no clear trigger. On each
run of the incremental parser we thus check the third situation for each uncommitted
language box in the program. Fortunately, unless the user has manually marked some
language boxes as uncommitted, the only uncommitted language boxes can be in
the location of the cursor, so their number is typically small and the performance
implications trivial.

\subsubsection{Resizing}

Although we do not change the starting position of an uncommitted language box
(which can be highly distracting), its end position can be automatically
changed to encompass more or less content (Figure~\ref{intro_example} shows an
example of the former). A language box is expanded to encompass more content
if: its parse tree does not contain a syntax error; if encompassing the
additional content does not cause a syntax error within the language box; and
if removing the content from the outer language does not cause syntax errors in
the first non-whitespace token in the outer language.
Language boxes are shrunk to encompass less content if they
contain a syntax error and moving the content to the outer language both
fixes the error inside the language box and does not
introduce additional syntax errors in the outer language.

Both growing and shrinking can be handled with our existing
candidates recogniser (see Section~\ref{sec:recognisers}).
While there is a clear indicator for when a language box may need to be shrunk
(e.g.~when it contains an error), this is not the case for expansion. We
must therefore run our altered candidates recogniser at the start of
each uncommitted language box; fortunately, as in automatic language box
removal (see Section~\ref{sec:removing}), there are almost never enough of
these to cause performance concerns. The candidates recogniser
returns all the possible right hand extents of the language box. We then
filter out candidates which do not meet the above conditions. If none remain,
the algorithm completes; if one remains,
we resize the language box appropriately; if more than one remains, we present
the multiple options to the user in the same way as in Section~\ref{multiple candidates}.

\subsection{Highly Ambiguous Compositions}
\label{sec:highly ambiguous compositions}

Some language compositions are so fundamentally ambiguous that normal automatic
language boxes do not work well. For example, consider a composition of Java
(or any other programming language!) with HTML, where HTML language boxes can
be used wherever Java expressions are valid. Since HTML's lexer can match
almost any text, nearly all syntax errors in Java can be resolved by wrapping text in
an HTML language box, which is unlikely to match the user's intentions.

To help with such cases, we thus have to relax the `no hints' constraint from
Section~\ref{the problem}, allowing language composition authors to specify
either the valid or the invalid token types which can appear at the start of
a language box. For example, in the case of the Java and HTML composition, a
good choice is to specify HTML tags as being the only valid starting token
types for a candidate language box. Any such hints given by language composition authors
are checked in the candidates recogniser so that inappropriate candidates do
not cause pointless work in later stages of the algorithm.

\section{Limitations}
\label{sec:lbox_limitations}

Since automatic language boxes are a default disambiguation mechanism,
there are inevitably situations where they do not perform as well as hoped.
Despite (as we shall see in Section~\ref{sec:evaluation}) such cases being
rare, it is useful to enumerate some of their fundamental weaknesses.

Although our solution to highly ambiguous compositions (see Section~\ref{sec:highly
ambiguous compositions}) works well when the outer language matches specific
input (e.g.~Java) and the inner language matches nearly anything (e.g.~HTML),
the reverse situation does not work well. For example, if we compose HTML and Java where
Java expressions are valid wherever HTML tags are valid, automatic language
boxes almost never trigger, since there are few ways of making a syntax error
in HTML. The only way this can be solved is by using a consideration
heuristic which is not triggered by syntax errors, but it is not clear to
us what a good heuristic along these lines might look like.

Lexical ambiguities between languages can also cause subtle problems. For
example, imagine that we compose Java and Lua, such that Lua expressions are
valid wherever Java expressions are valid and then input `\texttt{int x = 3 //
4 + 1;}'. This leads to a syntax error at the beginning of the line following
this statement and no automatic language boxes are inserted. However, this is
confusing for users, since `\texttt{//}' is Lua's integer division operator,
and they might reasonably expect a Lua language box to be put around `\texttt{3
// 4}' and/or `\texttt{3 // 4 + 1}'. However, since `\texttt{//}' is the Java
comment prefix, `\texttt{// 4 + 1;}' is ignored entirely, and the almost inevitable resulting
syntax error is postponed to the following line or later: in such cases, not even
the line candidates heuristic can find a starting point for language boxes that
matches a human's intuition. Although this example could be solved by searching
a fixed number of lines backwards, only an unbounded backwards
search, with its obvious performance problems, can solve the general case.
Fortunately, such ambiguities are sufficiently rare that maintaining the simplicity
of the line candidates heuristic seems a reasonable choice.

Automatic language boxes inherit incremental parsing's weaknesses on multiline
comments and strings. For example, in a naive Java grammar, typing
`\texttt{/*}' without the matching `\texttt{*/}' causes the entire rest of the
file (whether it is inside or outside a language box) to be relexed and tokens
flattened. Interestingly, a slight variant on language boxes solves this problem
for incremental parsing \cite[p.~108--122]{diekmann18editing}.
However, this is not applicable to our situation, where candidates
recognisers may end up lexing until the end of the file. Fortunately, this is
unlikely to be a performance problem in practice. First, this can only happen
if an inner language has sufficient lexical overlap with the outer language
(e.g.~sharing the same syntax for comments) and the rest of a file matches both
language's lexing rules.  Second, lexing is a fast activity in general and
particularly fast in the recogniser because we are not mutating the parse tree.

\section{Evaluation}
\label{sec:evaluation}

To evaluate the efficacy of automatic language boxes, a large-scale experiment
is necessary. We first present our methodology before looking at the results of
our experiment.

\subsection{Methodology}

Since there is no equivalent work that we know of, we have to
define our own methodology.

First, in order to produce numbers that are plausibly representative of situations
that real users might encounter, we created 12 language compositions of
real-world languages, and created \totalinsertions tests from real-world
programs. Each test is a tuple (\emph{base file}, \emph{base file function
definition or expression offset}, \emph{base file function definition or
expression span}, \emph{function definition or expression fragment}) where
`base file' is an instance of the outer language in the composition and
`fragment' is an instance of the inner language. For each test, we then loaded
\emph{base file} into our extension of Eco; emulated key presses which move the
cursor to the \emph{offset} and deleted \emph{span} characters; and then
emulated key presses which inserted \emph{fragment}. For each test we recorded
whether a language box was inserted or not, the number of characters covered
by any such language box, and the cause of any errors (e.g.~an insertion
led to subsequent errors; or no candidates were found at all).

Second, we need to classify the outcome of each test. The overall question we
want an answer to is: do automatic language boxes work well in most
cases? Answering this is not completely trivial, because there are several
possible outcomes from inserting text in an inner language. We
break these down into six categories: \label{detailed categories}

\begin{description*}
  \item[Complete insertion] A language box was automatically inserted around
    all of the fragment.
  \item[Partial insertion (no errors)] A language box was automatically
    inserted around part of the fragment, and the resulting text that was left
    in the outer language did not cause any errors.
  \item[Partial insertion (errors)] A language box was automatically inserted
    around part of the fragment, but the resulting text that was left
    in the outer language caused one or more syntax errors later in the file.
  \item[No insertion (valid)] No language box was automatically inserted because
    the fragment was valid in the outer language.
  \item[No insertion (errors)] No language box was automatically inserted even
    though the fragment was not valid in the outer language.
  \item[No insertion (multi)] No language box was automatically inserted because
    there were multiple candidates.
\end{description*}

We then group these under `acceptable' -- complete insertion, partial insertion
(no errors), no insertion (valid), and no insertion (multi) -- and
`unacceptable' -- partial insertion (errors), no insertion (errors).
If automatic language boxes are to be useful, we would need to see a high value
for the acceptable group and a low value for the unacceptable group.

\subsubsection{Language Compositions}

To create our 12 language compositions, we used the grammars of 4 real-world
languages (Java 5, Lua 5.3, PHP 5.6, and SQLite 3.27).
For each language composition of $L_1$ (outer) and $L_2$ (inner), we allow
expressions and function definitions in $L_2$ to be used wherever expressions
and function definitions in $L_1$ are normally valid. For example, in the
composition JavaLua, Java is the outer language and Lua the inner
language: Lua function definitions and expressions can be used wherever Java
function definitions and expressions (respectively) are valid. The one exception
to this is when SQL is the inner language: since our corpus consists of SQL
statements, it makes little sense to allow SQL statements where a Java/Lua/PHP
function is valid. In this case, we restrict the composition to only insert SQL
statements where Java/Lua/PHP expressions are valid (i.e.~we do not insert
function definitions at all). The
particular language compositions we created can be seen as the $x$-axis labels
in Table~\ref{tbl:valid}.

\subsubsection{Program Corpus}

We then collected a corpus of inputs in each language.
For PHP, we used the source code of Wordpress 4.6.13, which consists of \corpusphpfiles files that are
valid when parsed with our PHP grammar, which total \corpusphploc LoC after we
have stripped out the embedded HTML (which is not part of the PHP
language as such). For Java, we used the Java Standard Library 5 which consists
of \corpusjavafiles files and \corpusjavaloc LoC.
For Lua, we used Lua's test suite, which consists of \corpusluafiles files totalling \corpuslualoc LoC.
For SQL, we used SQLite's test suite, consisting of \corpussqlfiles test files which
contain \corpussqltests tests totalling \corpussqlloc LoC.

\begin{table*}[tb]
    \begin{tabular}{l  c  c  c  c  c  c  c  c  c  c  c  c  c }
    \toprule
        & \rotatebox{65}{JavaLua} & \rotatebox{65}{JavaPHP} & \rotatebox{65}{JavaSQL} & \rotatebox{65}{LuaJava} & \rotatebox{65}{LuaPHP} & \rotatebox{65}{LuaSQL} & \rotatebox{65}{PHPJava} & \rotatebox{65}{PHPLua} & \rotatebox{65}{PHPSQL} & \rotatebox{65}{SQLJava} & \rotatebox{65}{SQLLua} & \rotatebox{65}{SQLPHP} & \rotatebox{65}{Overall} \\
    \midrule
    \# Tests & 888 & 864 & 573 & 275 & 255 & 130 & 562 & 529 & 355 & 233 & 212 & 224 & 5,100 \\
    \midrule
    All & 98.8\% & 93.4\% & 99.3\% & 98.9\% & 94.9\% & 95.4\% & 99.3\% & 97.7\% & 97.5\% & 97.9\% & 97.6\% & 85.3\% & 96.8\% \\
    Parse tree & 76.4\% & 91.1\% & 100.0\% & 98.5\% & 93.3\% & 93.8\% & 99.5\% & 68.6\% & 98.3\% & 97.9\% & 96.2\% & 83.9\% & 89.4\% \\
    Stack & 96.6\% & 93.8\% & 99.3\% & 97.5\% & 92.9\% & 67.7\% & 99.5\% & 97.9\% & 97.5\% & 97.0\% & 97.6\% & 85.3\% & 95.6\% \\
    Line & 96.4\% & 79.9\% & 99.0\% & 98.9\% & 94.9\% & 95.4\% & 99.3\% & 95.5\% & 97.5\% & 97.9\% & 97.6\% & 85.3\% & 93.8\% \\
    \bottomrule
\end{tabular}

    \vspace{7pt}
    \caption{The total percentage of acceptable outcomes for each benchmark and
      heuristic. Acceptable outcomes are that: an automatic language box
      (covering all or part of the fragment) was inserted, causing no syntax
      errors; no language box was inserted since the fragment was also valid in
      the outer language; or there were multiple candidates which were presented to the
      user. Unacceptable outcomes are those which lead to syntax errors.}
    \label{tbl:valid}
    \vspace{-6pt}
\end{table*}

We then ran each file in the corpus through a parser to identify the offsets and spans
of function definitions and expressions, collecting the longest match when an
expression contained several subexpressions. For PHP and Java
we found that most expressions are so simple (e.g.~a function call with numeric
parameters) that they make highly repetitive, and uninformative, tests: in
these languages, we thus
extracted only expressions which form the right hand side of assignments, since
we found that these more often contain somewhat interesting expressions. The
situation in our Lua corpus was almost reversed, with many of the syntactically
interesting expressions occurring in \texttt{assert} statements. We thus
extracted all expressions from Lua files. Overall, we identified \corpusphpexprs
expressions and \corpusphpfuncs functions for PHP, \corpusjavaexprs expressions and \corpusjavafuncs functions for
Java, \corpusluaexprs expressions and \corpusluafuncs functions for Lua, and \corpussqlstmts statements for SQL.

\subsection{Results}

\begin{table*}[tb]
    \begin{tabular}{l  c  c  c  c  c  c  c }
    \toprule
        & \multicolumn{1}{p{2cm}}{\centering Complete insertion\\(No errors)} & \multicolumn{1}{p{2cm}}{\centering Partial insertion\\(No errors)} & \multicolumn{1}{p{2cm}}{\centering Partial insertion\\(Errors)} & \multicolumn{1}{p{2cm}}{\centering No insertion\\(Valid)} & \multicolumn{1}{p{2cm}}{\centering No insertion\\(Errors)} & \multicolumn{1}{p{2cm}}{\centering No insertion\\(Multi)} \\
    \midrule
    All & 65.8\% & 2.8\% & 2.3\% & 25.3\% & 0.8\% & 2.9\% \\
    Parse tree & 58.5\% & 3.1\% & 1.0\% & 25.3\% & 9.6\% & 2.5\% \\
    Stack & 64.9\% & 2.9\% & 2.0\% & 25.3\% & 2.4\% & 2.5\% \\
    Line & 63.3\% & 2.8\% & 2.3\% & 25.3\% & 3.8\% & 2.5\% \\
    \bottomrule
\end{tabular}

    \vspace{7pt}
    \caption{The total percentage of outcomes by category.
\ifarxiv
  See
    page~\pageref{detailed categories} for a detailed description of each category.
\fi
  The largest categories by far are `complete insertion (no errors)', which
  means that a language box was inserted around the entire input and `no
  insertion (valid)' which means that no language box was inserted because the
  fragment was valid in the outer language. Significantly, the unacceptable
  categories (`partial insertion (error)' and `no insertion (error)') constitute
  a very small portion of the total.}
    \label{tbl:breakdown}
    \vspace{-5pt}
\end{table*}

We break the results of our experiment down in two ways: by acceptable results
per language composition (Table~\ref{tbl:valid}); and by detailed outcome in the
overall benchmark suite (Table~\ref{tbl:breakdown}). In both cases we show
the differences in the various candidates heuristics, which shows that while
each has strengths and weaknesses, collectively they work well.
\ifarxiv
Appendix~\ref{apdx:tables}
contains several tables that give a more detailed breakdown of information
that is, in a sense, a combination of that between Tables~\ref{tbl:valid}
and~\ref{tbl:breakdown}.
\fi

The results of Table~\ref{tbl:valid} are clear: \validalloverall of tests have
an acceptable outcome with the `all' candidates heuristic. Table~\ref{tbl:breakdown} allows us to explore the
sub-categories. \breakdownallvalidsame of tests insert a complete language box
around the fragment (by definition making the program syntactically correct) and
\breakdownallnovalid of tests have fragments which are valid in the inner and
outer language. If we exclude tests whose fragments are valid in the inner and
outer language (i.e.~cases where the disambiguation mechanism cannot
expect to do anything useful), then \breakdownallexclude of the remaining cases
insert a language box around the whole fragment --
in other words, automatic language boxes match one reasonable (if slightly
naive) expectation in the vast majority of cases. The remainder of cases are
mostly acceptable outcomes: partial language box insertions without errors
(i.e.~part of the fragment is put in a language box but the remainder remains
in the outer language) and no insertions due to multiple candidates are both roughly
equal in proportion.

That leaves the two unacceptable cases. `Partial insertions (errors)'
(\breakdownallinvaliddiff of tests) are cases where a language box was inserted
around part of the fragment, but some was left in the outer language, causing a
syntax error later in the file. One solution to this might seem to be to
increase our search of the parse tree for errors after a language box has been
inserted, but this would come at a cost since we cannot reliably distinguish syntax
errors caused by language box insertion from syntax errors caused by
incomplete or deliberately incorrect user input. A better solution might be to
experiment with comparing the number and location of syntax errors before and
after candidate language box insertion, though this can only lessen, not fully
solve, the problem. `No insertions (errors)' (\breakdownallnoerror of tests) means that a language box was not
inserted around the fragment because it seemed to be valid in the outer
language even though it caused a syntax error later in the program.
Solving this problem suffers from the same
problem as for partial insertions with errors, and, since it is such a small
proportion, is not a significant issue.

Both Tables~\ref{tbl:valid} and \ref{tbl:breakdown} present the statistics for
our individual candidates heuristics. One might assume that the \texttt{lines}
candidates heuristic is the most effective, because it correctly deals with
nearly all insertions of expression fragments. However, its weakness is the
JavaPHP composition with only \validlinejavaphp tests having an acceptable
outcome. This is largely because the initial line of a PHP function
(e.g.~`\texttt{function x() \{}') looks like a Java function (with a return type
`\texttt{function}'): it is thus parsed as a Java function which leads to syntax
errors on subsequent lines of PHP code.

We expected the parse tree candidates
heuristic to be more effective than the stack candidates heuristic, but the
latter (\validstackoverall) has a noticeably greater proportion of acceptable
outcomes than the former (\validhistoverall). A generic weakness of the parse
tree heuristic is that many languages have grammar rules along the lines of
`\texttt{if (\emph{expr}) \{ ... \}}' i.e.~we expect a language box to be
inserted in the middle of the grammar rule. This defeats the parse tree
heuristic entirely, though the stack heuristic works as per human intuition.
The stack candidates heuristic, however,
performs poorly on LuaSQL because of the nature of Lua's grammar. For example,
in the mixed Lua/SQL program `\texttt{x = SELECT a, b FROM t}', everything up to
`\texttt{b}' can be parsed in Lua, leading to an LR reduction action which
removes `\texttt{SELECT}' from the parsing stack, making it impossible for the
heuristic to identify that as a valid location to insert a language box at.

\subsection{Threats to Validity}

There are two overall threats to validity to our evaluation.

First, it is possible that our suite of language compositions and corpus of
programs are unrepresentative. We have reduced these chances somewhat by using
4 real-world programming languages and a fairly large number of well-known
programs written in those languages, but it is possible to compose
very different styles of languages, and to compose them in very different ways.
Doing so might change our view of the automatic language boxes algorithm.

Second, we have made various suggestions about performance needs in this paper, but Eco is
a poor vehicle for evaluating whether automatic language boxes match such claims:
it is written in Python and the GUI runs only
in CPython, a particularly slow implementation; and Eco's data-structures
were originally designed to aid experimentation, and not performance. For example, nodes in Eco's
parse tree are very heavy weight (with a base size of around 2KiB per node, which grows
rapidly as undo history is added), and do not include any of the optimisations
described in~\citet{wagner98practicalalgorithms} (which would reduce the
base node size to approximately a tenth of its current size and substantially reduce the costs of
additional undo history). Despite that, editing performance with automatic
language boxes is rarely noticeable. To try and get some idea of what the
overhead might be with a more efficient implementation, we altered Eco so that
it can be run without the GUI on PyPy, a faster implementation of Python. On a
Xeon CPU E3-1270 3.60GHz machine, we then recorded the time each fragment took
to be inserted. The average per-keypress wall-clock time (including everything
associated with automatic language boxes) was 0.004s. Out of \totalinsertions tests,
7 (0.14\%) had an average per-keypress wall-clock time above 0.1s (often
considered the threshold at which humans start to perceive some lag from
typing), with a worst case of 0.30s. These results are fairly good as-is,
although we believe that a more efficient implementation could reduce these
timings by at least an order of magnitude.

\section{Conclusions}

In this paper we showed that default disambiguation within an
incremental parser works well for language composition, providing a new point
in the editing design space.

\begin{acks}
This research was funded by the EPSRC Fellowship \emph{Lecture} (EP/L02344X/1).
\end{acks}

\bibliography{bib}


\begin{thebibliography}{13}


\ifx \showCODEN    \undefined \def \showCODEN     #1{\unskip}     \fi
\ifx \showDOI      \undefined \def \showDOI       #1{#1}\fi
\ifx \showISBNx    \undefined \def \showISBNx     #1{\unskip}     \fi
\ifx \showISBNxiii \undefined \def \showISBNxiii  #1{\unskip}     \fi
\ifx \showISSN     \undefined \def \showISSN      #1{\unskip}     \fi
\ifx \showLCCN     \undefined \def \showLCCN      #1{\unskip}     \fi
\ifx \shownote     \undefined \def \shownote      #1{#1}          \fi
\ifx \showarticletitle \undefined \def \showarticletitle #1{#1}   \fi
\ifx \showURL      \undefined \def \showURL       {\relax}        \fi
\providecommand\bibfield[2]{#2}
\providecommand\bibinfo[2]{#2}
\providecommand\natexlab[1]{#1}
\providecommand\showeprint[2][]{arXiv:#2}

\bibitem[\protect\citeauthoryear{Bravenboer, Vermaas, Vinju, and
  Visser}{Bravenboer et~al\mbox{.}}{2005}]%
        {bravenboer05generalized}
\bibfield{author}{\bibinfo{person}{Martin Bravenboer}, \bibinfo{person}{Rob
  Vermaas}, \bibinfo{person}{Jurgen Vinju}, {and} \bibinfo{person}{Eelco
  Visser}.} \bibinfo{year}{2005}\natexlab{}.
\newblock \showarticletitle{Generalized type-based disambiguation of meta
  programs with concrete object syntax}. In \bibinfo{booktitle}{\emph{GPCE}}.
  \bibinfo{pages}{157--172}.
\newblock


\bibitem[\protect\citeauthoryear{Cantor}{Cantor}{1962}]%
        {cantor62ambiguity}
\bibfield{author}{\bibinfo{person}{David~G. Cantor}.}
  \bibinfo{year}{1962}\natexlab{}.
\newblock \showarticletitle{On the Ambiguity Problem of Backus Systems}.
\newblock \bibinfo{journal}{\emph{J.~ACM}} \bibinfo{volume}{9},
  \bibinfo{number}{4} (\bibinfo{date}{Oct.} \bibinfo{year}{1962}),
  \bibinfo{pages}{477--479}.
\newblock


\bibitem[\protect\citeauthoryear{Diekmann}{Diekmann}{2019}]%
        {diekmann18editing}
\bibfield{author}{\bibinfo{person}{Lukas Diekmann}.}
  \bibinfo{year}{2019}\natexlab{}.
\newblock \emph{\bibinfo{title}{Editing composed languages}}.
\newblock \bibinfo{thesistype}{Ph.D. Dissertation}. \bibinfo{school}{King's
  College London}.
\newblock


\bibitem[\protect\citeauthoryear{Diekmann and Tratt}{Diekmann and
  Tratt}{2014}]%
        {diekmann14eco}
\bibfield{author}{\bibinfo{person}{Lukas Diekmann} {and}
  \bibinfo{person}{Laurence Tratt}.} \bibinfo{year}{2014}\natexlab{}.
\newblock \showarticletitle{{Eco}: A language composition editor}. In
  \bibinfo{booktitle}{\emph{SLE}}. \bibinfo{pages}{82--101}.
\newblock


\bibitem[\protect\citeauthoryear{Khwaja and Urban}{Khwaja and Urban}{1993}]%
        {khwaja93syntax}
\bibfield{author}{\bibinfo{person}{Amir~Ali Khwaja} {and}
  \bibinfo{person}{Joseph~E. Urban}.} \bibinfo{year}{1993}\natexlab{}.
\newblock \showarticletitle{Syntax-directed Editing Environments: Issues and
  Features}. In \bibinfo{booktitle}{\emph{SAC}}. \bibinfo{pages}{230--237}.
\newblock


\bibitem[\protect\citeauthoryear{Pech, Shatalin, and Voelter}{Pech
  et~al\mbox{.}}{2013}]%
        {pech13mps}
\bibfield{author}{\bibinfo{person}{Vaclav Pech}, \bibinfo{person}{Alex
  Shatalin}, {and} \bibinfo{person}{Markus Voelter}.}
  \bibinfo{year}{2013}\natexlab{}.
\newblock \showarticletitle{{JetBrains} {MPS} As a Tool for Extending {Java}}.
  In \bibinfo{booktitle}{\emph{PPPJ}}. \bibinfo{pages}{165--168}.
\newblock


\bibitem[\protect\citeauthoryear{Tratt}{Tratt}{2008}]%
        {tratt08domainspecific}
\bibfield{author}{\bibinfo{person}{Laurence Tratt}.}
  \bibinfo{year}{2008}\natexlab{}.
\newblock \showarticletitle{Domain Specific Language Implementation via
  Compile-Time Meta-Programming}.
\newblock \bibinfo{journal}{\emph{TOPLAS}} \bibinfo{volume}{30},
  \bibinfo{number}{6} (\bibinfo{date}{Oct.} \bibinfo{year}{2008}),
  \bibinfo{pages}{1--40}.
\newblock


\bibitem[\protect\citeauthoryear{van Eijck}{van Eijck}{2007}]%
        {eijck07accept}
\bibfield{author}{\bibinfo{person}{Jan van Eijck}.}
  \bibinfo{year}{2007}\natexlab{}.
\newblock \showarticletitle{Let's Accept Rejects, But Only After Repairs}. In
  \bibinfo{booktitle}{\emph{Liber Amicorum for Paul Klint}}.
  \bibinfo{pages}{117--128}.
\newblock


\bibitem[\protect\citeauthoryear{Van~Wyk and Schwerdfeger}{Van~Wyk and
  Schwerdfeger}{2007}]%
        {wyk07context}
\bibfield{author}{\bibinfo{person}{Eric~R. Van~Wyk} {and}
  \bibinfo{person}{August~C. Schwerdfeger}.} \bibinfo{year}{2007}\natexlab{}.
\newblock \showarticletitle{Context-aware Scanning for Parsing Extensible
  Languages}. In \bibinfo{booktitle}{\emph{GPCE}}. \bibinfo{pages}{63--72}.
\newblock


\bibitem[\protect\citeauthoryear{Vasudevan and Tratt}{Vasudevan and
  Tratt}{2013}]%
        {vasudevan13detecting}
\bibfield{author}{\bibinfo{person}{Naveneetha Vasudevan} {and}
  \bibinfo{person}{Laurence Tratt}.} \bibinfo{year}{2013}\natexlab{}.
\newblock \showarticletitle{Detecting Ambiguity in Programming Language
  Grammars}. In \bibinfo{booktitle}{\emph{SLE}}. \bibinfo{pages}{157--176}.
\newblock


\bibitem[\protect\citeauthoryear{Vinju}{Vinju}{2005}]%
        {vinju05typedriven}
\bibfield{author}{\bibinfo{person}{Jurgen Vinju}.}
  \bibinfo{year}{2005}\natexlab{}.
\newblock \bibinfo{booktitle}{\emph{A Type-driven Approach to Concrete Meta
  Programming}}.
\newblock \bibinfo{type}{{T}echnical {R}eport} SEN-E0507.
  \bibinfo{institution}{CWI}.
\newblock


\bibitem[\protect\citeauthoryear{Visser et~al\mbox{.}}{Visser
  et~al\mbox{.}}{1997}]%
        {visser97scannerless}
\bibfield{author}{\bibinfo{person}{Eelco Visser} {et~al\mbox{.}}}
  \bibinfo{year}{1997}\natexlab{}.
\newblock \bibinfo{booktitle}{\emph{Scannerless generalized-LR parsing}}.
\newblock \bibinfo{type}{{T}echnical {R}eport} P9707.
  \bibinfo{institution}{Universiteit van Amsterdam}.
\newblock


\bibitem[\protect\citeauthoryear{Wagner}{Wagner}{1998}]%
        {wagner98practicalalgorithms}
\bibfield{author}{\bibinfo{person}{Tim~A. Wagner}.}
  \bibinfo{year}{1998}\natexlab{}.
\newblock \emph{\bibinfo{title}{Practical Algorithms for Incremental Software
  Development Environments}}.
\newblock \bibinfo{thesistype}{Ph.D. Dissertation}. \bibinfo{school}{University
  of California, Berkeley}.
\newblock


\end{thebibliography}

\ifarxiv
\appendix
\clearpage
\onecolumn

\section{Tables}
\label{apdx:tables}

In this appendix, we show detailed outcomes by candidates heuristic, in
similar fashion to Table~\ref{tbl:breakdown}.

\begin{table*}[h]
\begin{tabular}{l  c  c  c  c  c  c  c }
    \toprule
        & \multicolumn{1}{p{2cm}}{\centering Complete insertion\\(No errors)} & \multicolumn{1}{p{2cm}}{\centering Partial insertion\\(No errors)} & \multicolumn{1}{p{2cm}}{\centering Partial insertion\\(Errors)} & \multicolumn{1}{p{2cm}}{\centering No insertion\\(Valid)} & \multicolumn{1}{p{2cm}}{\centering No insertion\\(Errors)} & \multicolumn{1}{p{2cm}}{\centering No insertion\\(Multi)} \\
    \midrule
    JavaLua & 56.4\% & 2.4\% & 1.1\% & 34.6\% & 0.1\% & 5.4\% \\
    JavaPHP & 86.3\% & 4.7\% & 4.4\% & 1.5\% & 2.2\% & 0.8\% \\
    JavaSQL & 94.4\% & 0.2\% & 0.7\% & 4.7\% & 0.0\% & 0.0\% \\
    LuaJava & 51.3\% & 4.4\% & 0.4\% & 41.8\% & 0.7\% & 1.5\% \\
    LuaPHP & 84.3\% & 6.7\% & 1.2\% & 0.8\% & 3.9\% & 3.1\% \\
    LuaSQL & 86.2\% & 0.8\% & 1.5\% & 8.5\% & 3.1\% & 0.0\% \\
    PHPJava & 42.0\% & 4.8\% & 0.5\% & 49.6\% & 0.2\% & 2.8\% \\
    PHPLua & 47.8\% & 3.2\% & 1.7\% & 42.0\% & 0.4\% & 4.7\% \\
    PHPSQL & 93.5\% & 0.0\% & 2.3\% & 3.9\% & 0.3\% & 0.0\% \\
    SQLJava & 32.2\% & 0.9\% & 1.3\% & 60.9\% & 0.9\% & 3.9\% \\
    SQLLua & 32.1\% & 1.9\% & 2.4\% & 50.0\% & 0.0\% & 13.7\% \\
    SQLPHP & 60.3\% & 0.4\% & 14.7\% & 23.2\% & 0.0\% & 1.3\% \\
    \bottomrule
\end{tabular}

\caption{The \emph{all} candidates heuristics}
\end{table*}

\begin{table*}[h]
\begin{tabular}{l  c  c  c  c  c  c  c }
    \toprule
        & \multicolumn{1}{p{2cm}}{\centering Complete insertion\\(No errors)} & \multicolumn{1}{p{2cm}}{\centering Partial insertion\\(No errors)} & \multicolumn{1}{p{2cm}}{\centering Partial insertion\\(Errors)} & \multicolumn{1}{p{2cm}}{\centering No insertion\\(Valid)} & \multicolumn{1}{p{2cm}}{\centering No insertion\\(Errors)} & \multicolumn{1}{p{2cm}}{\centering No insertion\\(Multi)} \\
    \midrule
    JavaLua & 36.8\% & 0.0\% & 0.2\% & 34.6\% & 23.3\% & 5.0\% \\
    JavaPHP & 83.2\% & 6.4\% & 0.5\% & 1.5\% & 8.4\% & 0.0\% \\
    JavaSQL & 95.3\% & 0.0\% & 0.0\% & 4.7\% & 0.0\% & 0.0\% \\
    LuaJava & 52.7\% & 3.6\% & 0.4\% & 41.8\% & 1.1\% & 0.4\% \\
    LuaPHP & 82.7\% & 6.7\% & 0.8\% & 0.8\% & 5.9\% & 3.1\% \\
    LuaSQL & 84.6\% & 0.8\% & 1.5\% & 8.5\% & 4.6\% & 0.0\% \\
    PHPJava & 37.4\% & 10.7\% & 0.2\% & 49.8\% & 0.4\% & 1.6\% \\
    PHPLua & 19.7\% & 2.3\% & 0.2\% & 42.2\% & 31.2\% & 4.5\% \\
    PHPSQL & 94.4\% & 0.0\% & 1.4\% & 3.9\% & 0.3\% & 0.0\% \\
    SQLJava & 32.2\% & 0.4\% & 0.4\% & 60.9\% & 1.7\% & 4.3\% \\
    SQLLua & 32.5\% & 0.5\% & 1.9\% & 50.0\% & 1.9\% & 13.2\% \\
    SQLPHP & 58.5\% & 0.0\% & 12.5\% & 23.2\% & 3.6\% & 2.2\% \\
    \bottomrule
\end{tabular}

\caption{The \emph{parse tree} candidates heuristics}
\end{table*}

\begin{table*}[h]
\begin{tabular}{l  c  c  c  c  c  c  c }
    \toprule
        & \multicolumn{1}{p{2cm}}{\centering Complete insertion\\(No errors)} & \multicolumn{1}{p{2cm}}{\centering Partial insertion\\(No errors)} & \multicolumn{1}{p{2cm}}{\centering Partial insertion\\(Errors)} & \multicolumn{1}{p{2cm}}{\centering No insertion\\(Valid)} & \multicolumn{1}{p{2cm}}{\centering No insertion\\(Errors)} & \multicolumn{1}{p{2cm}}{\centering No insertion\\(Multi)} \\
    \midrule
    JavaLua & 54.5\% & 2.4\% & 0.5\% & 34.6\% & 2.8\% & 5.2\% \\
    JavaPHP & 87.3\% & 4.9\% & 4.1\% & 1.5\% & 2.2\% & 0.1\% \\
    JavaSQL & 94.4\% & 0.2\% & 0.7\% & 4.7\% & 0.0\% & 0.0\% \\
    LuaJava & 49.8\% & 5.1\% & 0.4\% & 41.8\% & 2.2\% & 0.7\% \\
    LuaPHP & 82.4\% & 6.7\% & 1.2\% & 0.8\% & 5.9\% & 3.1\% \\
    LuaSQL & 58.5\% & 0.8\% & 0.8\% & 8.5\% & 31.5\% & 0.0\% \\
    PHPJava & 43.4\% & 5.2\% & 0.4\% & 49.8\% & 0.2\% & 1.1\% \\
    PHPLua & 48.0\% & 3.4\% & 0.2\% & 42.2\% & 1.7\% & 4.3\% \\
    PHPSQL & 93.5\% & 0.0\% & 2.3\% & 3.9\% & 0.3\% & 0.0\% \\
    SQLJava & 31.3\% & 0.9\% & 1.7\% & 60.9\% & 1.3\% & 3.9\% \\
    SQLLua & 32.5\% & 1.9\% & 2.4\% & 50.0\% & 0.0\% & 13.2\% \\
    SQLPHP & 59.8\% & 0.4\% & 14.7\% & 23.2\% & 0.0\% & 1.8\% \\
    \bottomrule
\end{tabular}

\caption{The \emph{stack} candidates heuristics}
\end{table*}

\begin{table*}[h]
\begin{tabular}{l  c  c  c  c  c  c  c }
    \toprule
        & \multicolumn{1}{p{2cm}}{\centering Complete insertion\\(No errors)} & \multicolumn{1}{p{2cm}}{\centering Partial insertion\\(No errors)} & \multicolumn{1}{p{2cm}}{\centering Partial insertion\\(Errors)} & \multicolumn{1}{p{2cm}}{\centering No insertion\\(Valid)} & \multicolumn{1}{p{2cm}}{\centering No insertion\\(Errors)} & \multicolumn{1}{p{2cm}}{\centering No insertion\\(Multi)} \\
    \midrule
    JavaLua & 54.6\% & 2.4\% & 1.1\% & 34.6\% & 2.5\% & 4.8\% \\
    JavaPHP & 74.0\% & 4.2\% & 4.3\% & 1.5\% & 15.9\% & 0.2\% \\
    JavaSQL & 94.1\% & 0.2\% & 0.7\% & 4.7\% & 0.3\% & 0.0\% \\
    LuaJava & 51.3\% & 4.4\% & 0.4\% & 41.8\% & 0.7\% & 1.5\% \\
    LuaPHP & 84.3\% & 6.7\% & 1.2\% & 0.8\% & 3.9\% & 3.1\% \\
    LuaSQL & 86.2\% & 0.8\% & 1.5\% & 8.5\% & 3.1\% & 0.0\% \\
    PHPJava & 43.4\% & 5.2\% & 0.5\% & 49.6\% & 0.2\% & 1.1\% \\
    PHPLua & 45.6\% & 3.2\% & 1.7\% & 42.0\% & 2.6\% & 4.7\% \\
    PHPSQL & 93.5\% & 0.0\% & 2.3\% & 3.9\% & 0.3\% & 0.0\% \\
    SQLJava & 32.2\% & 0.9\% & 1.3\% & 60.9\% & 0.9\% & 3.9\% \\
    SQLLua & 32.1\% & 1.9\% & 2.4\% & 50.0\% & 0.0\% & 13.7\% \\
    SQLPHP & 60.3\% & 0.4\% & 14.7\% & 23.2\% & 0.0\% & 1.3\% \\
    \bottomrule
\end{tabular}

\caption{The \emph{line} candidates heuristics}
\end{table*}
\fi

\end{document}